\documentclass[11pt]{article}
\usepackage{graphicx}
\usepackage{dcolumn}
\usepackage{bm}
\usepackage{amsmath}   
\usepackage{graphics}
\usepackage{subfigure}

\begin{document}

\begin{center}

{\bf{\large Hadrons, Superconductor Vortices, and Cosmological Constant}}

\vspace{0.6cm}


{\bf  Keh-Fei Liu}

\end{center}

\begin{center}
\begin{flushleft}
{\it
Department of Physics and Astronomy, University of Kentucky, Lexington, Kentucky 40506, USA
}
\end{flushleft}
\end{center}

\begin{abstract}
     
  We explore the roles of the trace anomaly in several hadron properties. We derive the scale invariant expression for the pressure from the gravitational form factors (GFF) of QCD which results in consistent
results for the mass and rest energy from the GFF and those from the trace and the Hamiltonian of the  
energy-momentum tensor (EMT) operators. It is shown that the energy-equilibrium correspondence of hadrons infers an equation of state where the trace anomaly matrix element, emerging from the glue condensate in the vacuum, gives a negative constant pressure that leads to confinement, much like the confinement mechanism for the vortices in type II superconductors where the negative constant pressure is due to the cost of depleting the superconducting condensate.  We also note that both the trace anomaly in the QCD energy-momentum tensor and the cosmological constant in Einstein's equation are associated with the metric term which contributes to both energy and pressure.  Their difference in terms of the role the pressure plays is discussed. Finally, we note that a lattice calculation of the trace anomaly distribution in the pion has addressed a question about the trace anomaly contribution to the pion mass and suggests that there might be a connection between the conformal symmetry breaking and chiral symmetry breaking in this case.

\end{abstract}


\section{Introduction} \label{intro}

To discern the origin of the proton mass has been one of the major goals for the upcoming electron-ion collider (EIC)~\cite{Accardi:2012qut}. The quark masses have been well determined from experiments and lattice calculations -- the $u$ and $d$ quark masses are $m_u = 2.16^{+0.49}_{-0.26}$ MeV and $m_d = 4.67^{+0.48}_{-0.17}$ MeV in the 
$\overline{\rm{MS}}$ scheme at the renormalization scale of $\mu = 2$ Gev~\cite{ParticleDataGroup:2022pth}. These quark masses are only $\sim$ 1\% of the proton mass, this has raised the question  -- where does the rest of the proton mass come from?. A fair amount of work has been done to study the proton mass and energy decompositions recently in terms of the trace of the energy-momentum tensor matrix elements and the gravitational form factors~\cite{Ji:1994av,Ji:1995sv,Gao:2015aax,Teryaev:2016edw,Roberts:2016vyn,Lorce:2017xzd,Lorce:2018egm,Hatta:2018sqd,Metz:2020vxd,Kharzeev:2021qkd,Liu:2021gco,Burkert:2022hjz,Burkert:2023wzr}. In the meantime, the decomposition of the proton mass and energy has been studied on the lattice. A recent lattice calculation at the physical pion mass which takes into account the continuum limit extrapolation~\cite{Yang:2018nqn} shows that $\sim 91\%$ of the proton mass is due to the trace anomaly. The quark $\pi N$ sigma term and the strange sigma terms only contribute $\sim 9\%$ of the proton mass. On the other hand, the rest energy decomposition from the Hamiltonian involves the quark energy and the glue field energy in addition, while the trace anomaly contributes 1/4 of the rest energy in this case~\cite{Ji:1994av,Yang:2018nqn,Liu:2021gco}.

In this work, we shall explore the roles that the trace anomaly matrix elements play in several settings. From the gravitational form factors (GFF) of the energy-momentum tensor, it is found that there is a correspondence between the rest-energy decomposition and that of the equilibrium condition of the pressure, where the trace anomaly contributes with the same coefficient.  Since the trace anomaly in the hadron is due to the presence of a gluon condensate in the vacuum, this suggests that the trace anomaly contribution in the pressure is constant and negative which leads to hadron confinement. We shall point out that this is analogous to the vortices in type-II superconductors where the constant negative confining pressure is due to the depletion of Cooper pair condensate. 
We shall also discuss the similarities between the trace anomaly in QCD and the
cosmological constant. The different ways the pressure acts between the gauge theories and general relativity is also discussed. Finally, the spatial distribution of the trace anomaly in the pion resolves a pion mass question and, at the same time, suggests  that the conformal (scale) symmetry breaking maybe intricately linked to the spontaneous chiral symmetry breaking in the case of pion. 
We shall show that the trace anomaly form factor can be obtained from the moments of GPD which can be calculated on the lattice
as the gravitational form factors of the energy-momentum tensor (EMT) and the sigma terms which can also be evaluated on the lattice. This is the consequence of the conservation of the EMT. 

We shall start by discussing and reviewing the expression for the hadron mass and the decomposition of its rest energy in terms of the energy-momentum tensor in Sec.~\ref{mass}. Their expressions can also be obtained from the gravitational form factors as will be shown in Sec.~\ref{sec:GFF}. A rest energy-equilibrium correspondence which motivates an equation of state in volume and its implication on confinement is presented in Sec.~\ref{EEC}. The one-to-one correspondence between the rest energy
components of hadrons and the free energy of the vortices in type-II superconductors is illustrated in Sec.~\ref{vortices}. Their common confinement mechanism in terms of the condensates is pointed out. The analogies and differences between the trace anomaly matrix element and the cosmological constant are discussed in Sec.~\ref{cosmological_constant}. Finally, we discuss the relation between the spatial trace anomaly distribution in the pion and the behavior of the pion mass in Sec.~\ref{pion}. A summary and further discussion are given in Sec.~\ref{summary}. 
        
\section{Mass and Rest Energy} \label{mass}

Einstein's equation $E_0 = mc^2$ shows that the mass and the rest energy are equal. However, this does
not imply that their expressions in terms of their components or their origins are the same. In fact, many of their attributes are different. For one thing, the mass is a Lorentz scalar
while the energy is a component of the 4-momentum vector. In the example  of $e^+ e^-$ annihilation to two photons $e^+ e^- \longrightarrow \gamma\gamma$, it is pointed out that the mass of the two photon system deduced from the rest energy is $2 m_e$, not the sum of the two photon masses~\cite{Okun:1991nr,Okun:2000kf}.

The distinction between the mass and rest energy of hadrons in QCD can be demonstrated through the energy-momentum tensor (EMT). From the forward matrix element of the EMT
\begin{equation}
\langle P|T^{\mu\nu}|P\rangle = 2 P^{\mu}P^{\nu},
\end{equation}
the hadron mass can be obtained from the trace of the EMT. It is known that the trace of the EMT in QCD has an anomaly after renormalization~\cite{Chanowitz:1972vd,Crewther:1972kn,Chanowitz:1972da,Collins:1976yq}, 
\begin{equation} \label{trace}
T^{\mu}_{\mu} = T_{a\, \mu}^{\mu} + T_{\sigma\, \mu}^{\mu},
\end{equation}
where $T_{a\, \mu}^{\mu}$ is the trace anomaly and $T_{\sigma\, \mu}^{\mu}$ the sigma terms from the quark condensates. They have the expressions
\begin{eqnarray}  \label{trace_q,g}
T_{a\,\, \mu}^{\mu} = \frac{\beta(g)}{2g} G^{\alpha\beta} G_{\alpha\beta} 
\label{trace_g} + \sum_f \gamma_m (g)\, m_f \bar{\psi}_f \psi_f, \hspace{1cm}
T_{\sigma\, \mu}^{\mu} = \sum_f m_f \bar{\psi}_f \psi_f,  \label{trace_q}
\end{eqnarray}
and both are renormalization group invariant. Thus,
\begin{equation} \label{invariant_M}
M=\frac{\langle P| \int d^3  \vec{x}\, \gamma T^{\mu}_{\mu}(x)|P\rangle}{\langle P|P\rangle}
= \frac{\langle P| \int d^3  \vec{x}\, \gamma (T_{a\,\, \mu}^{\mu}(x) + T_{\sigma\, \mu}^{\mu}(x))|P\rangle}{\langle P|P\rangle}
\end{equation}
This shows that the hadron mass is both scale and frame independent,  just as expected for the mass which is a scalar.
For QCD with 2+1 light flavors, the $u$ and $d$ contributions to the $T_{\sigma\, \mu}^{\mu}$, i.e., the pion-nucleon sigma
term, is $\sigma_{\pi N} = \frac{m_u+m_d}{2} \langle P|\bar{u}u + \bar{d}d|P\rangle_{\vec{P} =0}/2M$ = 39.7 (3.6) MeV from the FLAG (Flavor Lattice Averaging Group) average~\cite{FlavourLatticeAveragingGroupFLAG:2021npn} and the strangeness sigma term is $\sigma_s = m_s \langle P|\bar{s}s |P\rangle_{\vec{P}=0}/2M$ = 40.2(3.9) MeV from a lattice calculation~\cite{Yang:2015uis}. Together, they account for \mbox{$\sim$ 8.5(8)\%} of the nucleon mass. The rest of the nucleon mass is due to the trace anomaly. 

On the other hand, the decomposition of the rest energy can be obtained from the Hamiltonian or the gravitational form factor (GFF). The Belinfante form of the EMT is a symmetric rank two tensor, 
which can be separated into traceless and trace components in irreducible representations~\cite{Ji:1994av,Ji:1995sv}
\begin{equation} \label{trace_separation}
T^{\mu\nu} = \overline{T}^{\mu\nu} + \frac{1}{4} g^{\mu\nu} T_{\rho}^{\rho}.
\end{equation}
 So far, this separation is
scale and scheme independent. $\bar{T}^{\mu\nu}$ can be further split into the quark and glue parts. In this case,
the Hamiltonian, being the spatial integral of $T^{00}$, i.e., $ H = \int d^3\vec{x}\,\, T^{00}(x)$, can be written as~\cite{Ji:1994av,Ji:1995sv}
\begin{equation}  \label{H_3-term}
H = H_q + H_g + \frac{1}{4} (H_{a} + H_{m}),
\end{equation}
where 
 \begin{eqnarray}
 H_q &=& \int d^3\vec{x}\, (\frac{i}{4} \sum_f \bar{\psi}_f \gamma^{\{0}\!\stackrel{\leftrightarrow}{D}\!{}^{0\}}\psi_f 
 - \frac{1}{4} T_{q\, \mu}^{\mu}),  \label{H_q}\\
 H_g &=& \int d^3\vec{x}\, \frac{1}{2} (B^2 + E^2),  \label{H_g}, \label{H_g} \\
 H_{a} &=&  \int d^3\vec{x}\,  T_{a\,\, \mu}^{\mu} = \frac{\beta(g)}{2g} G^{\alpha\beta} G_{\alpha\beta} 
\label{trace_g} + \sum_f  \gamma_m (g) \, m_f\bar{\psi}_f \psi_f \\
H_m &=& \int d^3\vec{x}\,  T_{q\,\, \mu}^{\mu} = \sum_f m_f \bar{\psi}_f \psi_f.  \label{H_m}
\end{eqnarray}  
$H_q$ and $H_g$ are the Hamiltonian operators for the quark energy and glue field energy.
They are in the form of bare operators with the understanding that
they need to be renormalized in a scheme in order to obtain the results for the
rest energy at certain scale. A practical and physical scheme to computer their matrix elements and compare with experiments is the lattice approach, where the bare operators are discretized, renormalized and mixed in the non-perturbative RI/MOM scheme and matched to the $\overline{\rm MS}$ scheme at 2 GeV~\cite{Yang:2018nqn}. The traceless operators in Eqs.~(\ref{H_q}) and (\ref{H_g}) are calculated with $T^{0i}$ or $T^{00} - T^{ii}/3$ operators~\cite{Yang:2018nqn,Constantinou:2020hdm,Wang:2021vqy}. 
On the other hand, $H_a$ and $H_m$ from the trace part of $T^{00}$ are the renormalization group invariant operators from the trace anomaly and the sigma terms as shown in Eq.~(\ref{trace_q,g}).
The $\sigma$ terms are calculated with direct calculations or through the Feynman-Hellman theorem~\cite{FlavourLatticeAveragingGroupFLAG:2021npn} on the lattice. 
It is shown that the trace anomaly emerges with the lattice regulation after renormalization~\cite{Caracciolo:1989pt,Makino:2014taa,DallaBrida:2020gux}. The only complication in the lattice calculation is that it can mix with the lower-dimensional operator $\bar{\psi}\psi$ with a $1/a$ power divergence. The trace anomaly matrix element has been calculated for the nucleon where the 
$\beta/2g$ and $\gamma_m$ are fitted with the pseudoscalar and vector meson mass relations in Eq. (\ref{invariant_M}) for a valence quark mass and checked for consistency with several valence quark masses~\cite{He:2021bof}. They can be used for other hadrons on the same lattice. The $1/a$ term has been included in the fit, but the signal is too weak to be isolated in this calculation at the lattice spacing 
$a = 0.114$ fm~\cite{He2022a}. We should note that
the traceless operators $H_q$ and $H_g$ do not mix with the trace operators $H_a$ and $H_m$ on the lattice. They are in different irreducible representations of the $O(4)$ group. 

The traceless EMT $\overline{T}^{00}$ matrix elements of the nucleon at rest are 3/4 of the second moments of the parton distribution functions (PDFs)~\cite{Ji:1994av,Ji:1995sv,Yang:2018nqn} which are the momentum fractions on the light front, i.e., $\langle x\rangle_q(\mu)$ and 
$\langle x\rangle_g(\mu)$. They incorporate renormalization and mixing of the operators of $H_q$ and $H_g$ in Eqs. ~(\ref{H_q}) and (\ref{H_g}) at the scale $\mu$. Thus, the rest energy for the proton with $2+1$ flavors is
\begin{equation} \label{E0H}
E_0 = \frac{3}{4} \,[\langle x\rangle_q (\mu)+ \langle x\rangle_g (\mu)] M + \frac{1}{4} [\langle H_a\rangle + \sigma_{\pi N} + \sigma_s].
\end{equation}
We shall define $\langle H_{...}\rangle = \langle P|H_{...}|P\rangle/\langle P|P\rangle$ at $\vec{P} = 0$.
We see that 3/4 of $E_0$ is due to the quark and glue momentum fractions (or quark and glue field energies) and 1/4 is from the trace anomaly and the $\sigma$ terms~\cite{Ji:1994av,Ji:1995sv}. 

\begin{figure}[htbp]     \centering
\subfigure[]
{\includegraphics[width=0.445\hsize]{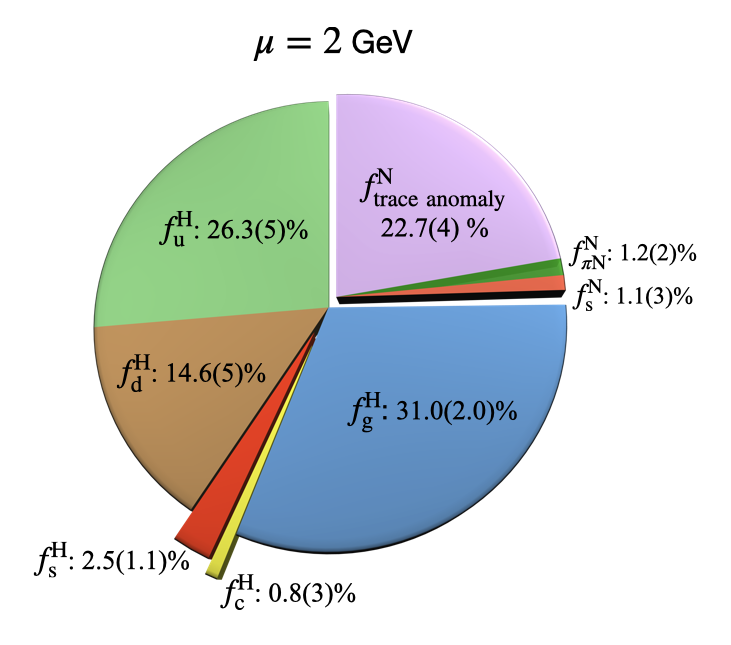}
  \label{RE-3f}}
\subfigure[]
{\raisebox{4ex}
{\includegraphics[width=0.465\hsize]{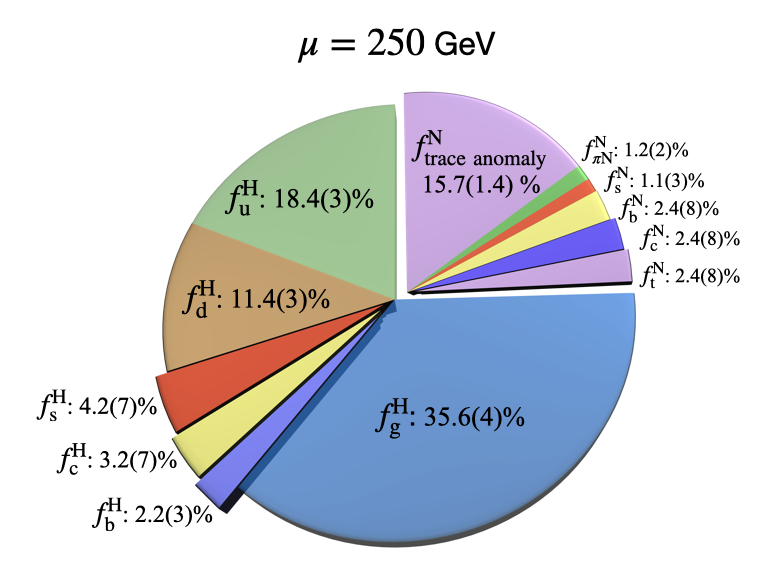}}
  \label{RE-6f}}
\caption{Proton rest energy decomposition in terms of quark sigma terms of different flavors, the trace anomaly and the quark and
glue momentum fractions. They are plotted as the percentage fractions of the proton mass. The fractions $f_{f,g}^H$ are from the second moments (3/4 $\langle x \rangle_{f,g}$) of the PDFs from the CT18 global analysis~~\cite{Hou:2019efy,T.J.Hou}. The sigma terms are obtained from lattice  calculations~\cite{Yang:2015uis,Gong:2013vja}. (a) is for the 2+1 flavor case at the scale of
$\mu = 2$ GeV and (b) for the case including the charm, bottom and top momentum fractions and their sigma terms
at the weak interaction scale  $\mu = 250$ GeV. \label{rest_energy}
}
  \end{figure}

We plot the fractional contributions of these components in the pie charts for the nucleon in Fig.~\ref{rest_energy}. 
The fractions in the figures include $f_{f,g}^{H} = \langle H_{f,g}\rangle/M = 3/4\,\langle x\rangle_{f,g}$, where the quark flavor $f = u+d\, (\pi N),s,c,b,t$, $f^N_{\rm trace\,\,anomaly} =  \langle H_a\rangle$/4M,  and 
$f_f^N =  \langle H_{m_f}\rangle$/4M.
For the fractions of the traceless part $f_{f,g}^H$, we shall use the second moments of the PDFs from 
CT18~\cite{Hou:2019efy,T.J.Hou} in the global analysis of experiments. The $\sigma$ terms are from the lattice calculations~\cite{Yang:2015uis,Gong:2013vja}  using the overlap fermions.   Since the separation of the quark and glue
momentum fractions are scale dependent, we plot the fractions at the hadronic scale at $\mu = 2 $ GeV in Fig.~\ref{RE-3f}
and considered $2+1$ flavors for the $\sigma$ terms. In Fig.~\ref{RE-6f}, we plot them at the weak scale of 
$\mu = 250$ GeV with 6 flavors for the $\sigma$ terms. As we see from Fig.~\ref{RE-3f} and Fig.~\ref{RE-6f}, when the scale is increased, the valence partons (i.e., $u$ and $d$) fractions are shifted more toward sea and gluon partons, but their total
contributions stays at 3/4 of the proton rest energy. Similarly, in the 1/4 contribution from the trace, the inclusion of heavy quarks ($c,b,$ and $t$) shrinks the trace anomaly contribution. This reflects the fact that, to leading order in the heavy quark expansion, the matrix element for the heavy quark $\bar{Q}Q$ is proportional to the glue $G^{\alpha\beta}G_{\alpha\beta}$ matrix element in the nucleon~\cite{Shifman:1978zn}, i.e., $m_Q \langle P | \bar{Q} Q | P \rangle_{\stackrel{\longrightarrow}{m_Q \rightarrow \infty}} - \frac{1}{3}(\frac{\alpha_s}{4 \pi})\langle  P | G^2 |P \rangle$. The coefficient $- \frac{1}{3}(\frac{\alpha_s}{4 \pi})$ corresponds to  the $n_f$ term in the leading $\alpha_s/4\pi$ expansion of $\frac{\beta(g)}{2g}$  with
a negative sign. This shows that for $n_f$ heavy enough quarks, the introduction of their sigma terms is largely absorbed by
the trace anomaly with a change in the $\beta$ function. The net contribution of
a heavy quark with mass $M_H$ is $\mathcal{O} (1/M_H)$, in accordance with the decoupling theorem~\cite{Appelquist:1974tg,Kaplan:1988ku}.  To study the quark mass dependence, a lattice
calculation with the overlap fermion has been carried out~\cite{Gong:2013vja}. It is found that the sigma terms for quark masses heavier than $\sim 1/2$ of the charm mass are the same within errors. We take this finding to mean that the sigma terms for the charm, beauty and top quarks are about the same. For the charm, it is found~\cite{Gong:2013vja} that $f_c^N = 0.024(8)$, which is taken to be the same for $f_b^N$ and $f_t^N$. 

Not all the components in the rest energy are associated with experimental observables so far, notably the $\sigma$ terms other than the $\pi N$ sigma term and the trace anomaly matrix element. However, all of these components are amenable to lattice calculations and they have been calculated on the lattice. 

   From the above discussion of the mass and rest energy, we realize that the question `Since the $u$ and $d$ quark masses are only $\sim 1$\% of the proton mass, where does the rest of proton mass come from?' is misguided. As we see from the above $e^+ e^- \rightarrow \gamma\gamma$ example and
the bound states like the hydrogen atom, the mass is not the sum of their constituent masses. If one attempts to separate the quark and the glue contributions in the mass expression in Eq.~\ref{trace}, they become scheme and scale dependent~\cite{Hatta:2018sqd,Metz:2020vxd}. Instead, one should ask `Since the nucleon sigma terms are small ($\sim 2$\% of the proton rest energy for the 2+1 flavor case), where does the rest of the nucleon rest energy come from?'.

\subsection{Gravitational Form Factors}   \label{sec:GFF}

The nucleon mass and rest energy can also be obtained from the gravitational form factors (GFF) of the EMT. 
They contain the following terms~\cite{Kobzarev:1962wt,Pagels:1966zza,Ji:1996ek}
for the quarks and gluons 
\begin{eqnarray} \label{GFF}
\langle P'| T_{q, g}^{\mu\nu}|P\rangle /2 M&=& \bar{u}(P')[A_{q,g}(q^2,\mu) \gamma^{(\mu} \bar{P}^{\nu)} +
B_{q,g}(q^2,\mu) \frac{\bar{P}^{(\mu} i \sigma^{\nu)\alpha} q_{\alpha}}{2M}   \nonumber \\
     &+&  D_{q,g}(q^2,\mu)\frac{q^{\mu}q^{\nu} - g^{\mu\nu}q^2}{M} + \bar{C}_{q,g}(q^2, \mu) M g^{\mu\nu} ] u(P)
\end{eqnarray}
where the forward matrix element $A_{q,g}(0)= \langle x\rangle_{q,g} (\mu)$ is the momentum fraction and $A_{q,g} (0) + B_{q,g} (0) = 2J_{q,g} (\mu)$ the angular momentum fraction~\cite{Ji:1996ek}. 
%
%
By making a connection to the stress tensor of the continuous medium, it is suggested~\cite{Polyakov:2002yz,Polyakov:2018zvc} that $D_{q,g}(q^2)$ is related to the internal force of the hadron and
encodes the shear forces and pressure distributions of the quarks and glue in the nucleon. Similarly, 
$\bar{C}(0)$ is suggested to be the pressure-volume work~\cite{Lorce:2017xzd,Lorce:2018egm}.  This identification is confirmed by realizing that it is equal to the normal stress $T^{ii}(0)$~\cite{Liu:2021gco}. We shall give a derivation of $\bar{C}$ in its
scale-invariant form and demonstrate that it plays a crucial role in obtaining consistent 
mass and rest energy decompositions in Eqs.~(\ref{invariant_M}) and (\ref{E0H}) from the irreducible representation of the EMT and those from the GFF.  More importantly, we shall explore its physical meaning and consequences in this work.

From the GFF in Eq.~(\ref{GFF}), one can get the total forward $\langle T_{\mu}^{\mu}\rangle$ and $\langle T^{00}\rangle$ matrix elements
\begin{equation}  \label{Tmumu}
\langle T_{\mu}^{\mu}\rangle = (\langle x\rangle_q (\mu)+ \langle x\rangle_g (\mu)) M + 4 (\bar{C}_q(0) + \bar{C}_g (0)) M,
\end{equation} 
\begin{equation}  \label{T00}
\langle T^{00}\rangle = (\langle x\rangle_q (\mu)+ \langle x\rangle_g (\mu)) M +  (\bar{C}_q(0) + \bar{C}_g (0)) M
\end{equation}
and $\bar{C}(0)$ is related to the the forward matrix element of the normal stress, i.e., $\langle T^{ii}\rangle$
\begin{equation} \label{Tii}
\langle T^{ii} \rangle =  - 3 \,(\bar{C}_q(0) + \bar{C}_g (0))M 
\end{equation}
Since $T^{ii} = T^{00} - T_{\mu}^{\mu}$, Eqs.~(\ref{trace_q,g}), (\ref{T00}) and (\ref{Tii}) can be combined to solve $\bar{C}(0)$ which gives
\begin{equation}  \label{barC}
\bar{C}_q(0) + \bar{C}_g (0) = \frac{1}{4} (f_a^N +  f_q^N - (\langle x\rangle_q (\mu) 
+ \langle x\rangle_g (\mu)))
\end{equation}
where $f_a^N$ and $f_q^N = \sum_f f_f^N$ are the fractions of trace anomaly and sigma term contributions to the nucleon mass and $\langle x\rangle_q (\mu)= \sum_f \langle x\rangle_i (\mu)$ is the total momentum fraction of the quarks. Note that $\bar{C}_q(0,\mu)$ and $\bar{C}_g(0,\mu)$ separately have $\mu$ dependence, but when combined in Eq.~(\ref{barC}), do not have scale dependence. From this point on, we drop the $\mu$ dependence in
$\langle x\rangle_q$ and $\langle x\rangle_g$ for simplicity. 
From the expression of $\bar{C}(0)$, one obtains from Eqs.~(\ref{Tmumu}) and ~(\ref{T00})
\begin{eqnarray}  \label{Tmumu00}
\langle T_{\mu}^{\mu}\rangle \! \!\! &=&\!\!\! M = (f_a^N + f_q^N) M, \nonumber \\
\langle T^{00}\rangle \!\!\!&=& \!\!\! E_0 = 3/4  (\langle x\rangle_q + \langle x\rangle_g) M + 1/4 (f_a^N + f_q^N) M.
\end{eqnarray}
which are in agreement with the mass expression from the trace in Eq.~(\ref{invariant_M}) and the rest energy decomposition from the Hamiltonian in Eq.~(\ref{E0H}) as a cross check. This is easy to understand. The GFFs are organized by Lorentz covariance and CPT symmetry, not by the irreducible representations of the EMT. As such, the $A_q$ and $A_g$ terms include both the traceless and
trace contributions. By subtracting the trace contributions in the $\bar{C}$ term in Eq.~(\ref{barC}), does one obtain the traceless contribution of $E_0$ from $\langle T^{00}\rangle$, which is $3/4  (\langle x\rangle_q + \langle x\rangle_g) M$ and the trace part is the remainder of $\bar{C}$, which is  $1/4 (f_a^N + f_q^N) M$ as shown in Eq.~(\ref{Tmumu00}). By the same token,  the $A_q$ and $A_g$ terms in the trace matrix element $\langle T_{\mu}^{\mu}\rangle$ are cancelled by the same from the $\bar{C}$ so that the traceless part vanishes and it leaves the remainder of $\bar{C}$ to be just the trace $(f_a^N + f_q^N) M$ as  given in Eq.~(\ref{Tmumu00}). 

Since the EMT is conserved, i.e.,
$\partial_{\nu} T^{\mu \nu}=  0$, this leads to the sum $\bar{C}_q(0) + \bar{C}_g (0) = 0$ as can be readily verified
in Eq.~(\ref{barC}). In view of this, $\bar{C}$ has been dropped from the GFF in some of the recent
literature. As a consequence, one finds that 
$\langle T_{\mu}^{\mu}\rangle = \langle T^{00}\rangle = (\langle x\rangle_q + \langle x\rangle_g) M$. This is not a
correct physical decomposition, as $\langle T_{\mu}^{\mu}\rangle$ does not contain the trace anomaly contribution and the
rest energy from $\langle T^{00}\rangle$ does not have all the relevant physical contents as in Eq.~(\ref{E0H}). This will be explained in more detail in Sec.~\ref{EEC}. 

    As we mentioned, $\bar{C}(0)$ is the negative of the normal stress $\langle T^{ii}(0)\rangle$ in Eq.~(\ref{Tii}). Since the normal stress density is the pressure~\cite{Landau:1975pou}, $\langle T^{ii}(0)\rangle$ is the pressure-volume work.  Since the total $\bar{C}$ is zero, it is the equilibrium condition, i.e.,
\begin{equation}  \label{equilibrium}
PV = \frac{\langle T^{ii}(0)\rangle}{3} = - ( \bar{C}_q(0) + \bar{C}_g (0)) M = - \frac{1}{4} (f_a^N +  f_q^N) M  + \frac{1}{4} (\langle x\rangle_q + \langle x\rangle_g)) M = 0,
\end{equation}
where the positive pressure from the quarks' energies and the glue field energy are balanced by the negative pressure from the trace anomaly matrix element and quark $\sigma$ terms. In statistical mechanics,
the free energy $F = - kT \ln Z$ where $Z$ is the partition function in the canonical ensemble and the pressure $P = -\frac{\partial F}{\partial V}$. The canonical ensemble in QCD has been formulated in the path-integral formalism and lattice calculations have been performed at different baryon numbers to search for the critical point at finite density and temperature~\cite{Liu:2002qr,Alexandru:2005ix,Li:2011ee}. At zero temperature, $P = - \frac{\partial E}{\partial V}$. Consequently, it has implication on the volume dependence of the rest-energy~\cite{Liu:2021gco}. It is important to note that the equilibrium condition in Eq.~(\ref{equilibrium})  and the rest energy in Eq.~(\ref{E0H}) (same as in Eq.~(\ref{Tmumu00})) involve the 
same matrix elements, but with different coefficients. Since the same matrix elements are involved and
$PV = - \frac{\partial E}{\partial V} V$, the ratios of the coefficients of those terms in Eq.~(\ref{equilibrium}) to their counterparts in the rest-energy in Eq.~(\ref{Tmumu00})) are the negative exponents of the volume dependence in the equation of state. In particular, the negative unit ratio for the trace terms indicates that their contributions to the rest energy is linear in volume and, thus, yields a negative constant pressure which confines the hadrons.  Whereas, the positive pressure-volume work is $ - \frac{1}{3}$ of the quark and glue energies in $E_0$. This infers that their volume dependence is $V^{-1/3}$. Further examination of the significance of this rest energy - equilibrium correspondence are presented in the next section, Sec.~\ref{EEC}.

\subsection{Energy -- equilibrium correspondence and equation of state} \label{EEC}


        The energy-equilibrium correspondence derived from the GFF in the above section should be a general feature in the bound or confined states which have characteristic sizes. The rest energies can in principle
be expressed in terms of their sizes, i.e., equations of state $E (V)$. For a stable state, it would inevitably involve at least two types of energies with different size dependences so that there can be an equilibrium size with the condition 
$\frac{d E}{d V}|_{V_0} = 0$ and the stability condition $\frac{d^2 E}{d^2 V}|_{V_0} > 0$. In particular, when these energies
with power dependence on the volume (or radius), 
\begin{equation}
E (V) = \sum_i \epsilon_i V^{p_i},
\end{equation}
then the equilibrium condition is
\begin{equation}   \label{power}
PV = - \frac{d E}{dV} V = - \sum_i p_i\, \epsilon_i V^{p_i}|_{V_0} = 0.
\end{equation}
It involves the same $\epsilon_i V^{p_i}$ terms as in $E(V)$, but weighted with the power of the volume dependence
$p_i$ and a negative sign.

     The rest energy -- equilibrium correspondence can be demonstrated in a few examples of models which have explicit equations of state.

\begin{enumerate}

\item MIT bag model:

       In this model with relativistic quarks and gluons confined in a bag with certain bag boundary conditions~\cite{Chodos:1974je,Chodos:1974pn},
the equation of state of a hadron is expressed as
\begin{equation}
E(V) = BV + \frac{\Sigma_{q,g}}{R}
\end{equation}
where $B$ is the confining bag constant and $\frac{\Sigma_{q,g}}{R}$ are the eigenenergies of the quarks and
gluons with the boundary condition. The equilibrium radius $R$ is determined from $\frac{d E(V)}{dV}|_{V_0}= 0$ and
the PV equilibrium is
\begin{equation} \label{bagPV}
PV = - \frac{dE}{dV} V|_{V_0} = - (BV_0 - \frac{1}{3} \frac{\Sigma_{q,g}}{R_0}) =0,
\end{equation}
where $R_0 = (3 V_0/4 \pi)^{1/3}$. The rest energy is then $E_0 = BV_0 + \frac{\Sigma_{q,g}}{R_0}$.
The unit and $ - \frac{1}{3}$ factors in Eq.~(\ref{bagPV}) in front of the the $BV_0$ and $\frac{\Sigma_{q,g}}{R_0}$ terms  simply reflect their volume dependences in $E (V)$ as demonstrated in Eq.~(\ref{power}). 

\item Non-relativistic potentials:

For one-body non-relativistic potential problems with a potential as a power of the radius, i.e., $V(r) = k r^n$,
the rest energy is the sum of the kinetic energy and the potential energy
\begin{equation}  \label{T+V}
E_0 = \langle T\rangle + \langle V\rangle.
\end{equation}
With a variational wavefunction characterized by its size $R$, the kinetic energy would scale as $R^{-2}$ and
the potential energy scales as $R^n$. Therefore the equation of state for $E(R)$ is
\begin{equation} \label{NRE}
E(R) = \frac{\epsilon_T}{R^2} + \epsilon_V R^n,
\end{equation}
where $\epsilon_T = \langle T\rangle (R) R^2$ and $\epsilon_V = \langle V\rangle (R)/ R^n$ are constant matrix elements. 
Upon differentiation, one obtains
\begin{equation}  \label{FR}
FR = - R\, dE/dR|_{R_0} = 2 \langle T\rangle (R_0)- n \langle V\rangle (R_0)= 0
\end{equation}
To the extent that the matrix elements $\langle T\rangle (R_0)$ and $\langle V\rangle (R_0)$ from the variational approach are  good approximation of those from the  solution of the Schr\"{o}dinger equation,  Eq.~(\ref{FR}) is just the
 well-known virial theorem. Thus, the virial theorm for the bound states with the non-relativisitc potential 
 $r^n$ can be understood in terms of the equilibrium condition. Again, the factors in front of the scaled matrix elements
in the equilibrium condition in Eq.~(\ref{FR})  reflect the exponents of the size dependence of $E(R)$ in Eq.~(\ref{NRE}).

\item Quantum many-body problem:

         It is proven in a general many-body problem with two-body potential~\cite{MS2010} that the pressure-volume work from the normal stress is
\begin{equation}
3 PV = T_{ii} = 2 \sum_i \Biggl \langle \frac{\vec{p_i}^2}{2m_i} \Biggr\rangle - \frac{1}{2} \sum_{i \neq j} \langle 
|\vec{r}_i - \vec{r}_j| V_{ij}' (|\vec{r}_i - \vec{r}_j|)\rangle.
\end{equation}
This is the Clausius form of the virial theorem.  The same virial expression is obtained from 
statistical mechanics~\cite{MS2010}  where the pressure is obtained from
\begin{equation}
P = kT \Biggl (\frac{\partial \ln Z}{\partial V} \Biggr ) = - \Biggl \langle \frac{\partial H}{\partial V}\Biggr \rangle,
\end{equation}
where $H$ is the Hamiltonian. 

For a bound state, the pressure-volume work vanishes (i.e., $PV = 0$) due to the cancellation between the kinetic energy and the potential energy contributions which is the equilibrium condition we discussed above.  To this end, it is physical and more meaningful to have both the kinetic and potential energies present in the decomposition of the rest energy as given in the above
examples. On the other hand, from the virial theorem, one can obtain for the case of the Coulomb potential,
$E_0 = - \langle T\rangle$ or $E_0 = \langle V\rangle/2$. But neither is an acceptable physical interpretation of the bound state energy decomposition of the hydrogen atom, as they do not have the full physical contents as in Eq.~(\ref{T+V}).
Consequently, expressing the hadron rest energy as $E_0 = \langle T^{00}\rangle = (\langle x\rangle_q + \langle x\rangle_g) M$
by dropping the $\bar{C}$ terms first in the GFF as discussed in Sec.~\ref{sec:GFF}, is not a 
valid physical decomposition. It lacks the `potential energy' from the trace part of the EMT. 

\end{enumerate}

From Eq.~(\ref{power}) and the above examples, it is clear that the rest-energy - equilibrium correspondence
between the rest energy in Eqs.~(\ref{E0H}), (\ref{Tmumu00}) and the $PV =0$ in Eq.~(\ref{equilibrium}) infers an equation of state for the hadron rest energy $E_H(V)$ as a function of $V$
\begin{equation}  \label{EoS}
E_H (V) = E_S + E_T = \epsilon_S V + \epsilon_T V^{-1/3}
\end{equation}
 where $\epsilon_S = E_S/V$ is the density of the singlet trace energy $E_S = 1/4 (f_a^N + f_q^N) M$ and
 $\epsilon_T = E_T V^{-1/3}$ is the scaled density of the triplet traceless energy 
 $E_T = 3/4 (\langle x\rangle_q + \langle x\rangle_g) M$. Incidentally, Eq.~(\ref{EoS}) has the same volume dependence
 as in the MIT bag model and the bag constant $B$ has been suggested to be from the gluon condensate~\cite{Jacobs:2004qv}. It can be identified as the QCD trace anomaly matrix element from the present study.

 The equilibrium condition $PV = - (E_S  - \frac{1}{3} E_T) = 0$ 
 corresponds to  Eq.~(\ref{equilibrium}). Thus, the emerged trace anomaly is the negative of the energy density of the gluon  condensate times the volume of the hadron and, thereby, the volume can be defined as
 \begin{equation} \label{volume}
 V = \frac{\int d^3 r \, \rho_{a} (r)}{|\langle \frac{\beta}{2g} \,G^{\alpha\beta}G_{\alpha\beta}\rangle|},
  \end{equation}
 where $\rho_{a} (r)$ is the radial distribution of the glue part of the trace anomaly in the hadron which can
 be obtained from the Fourier transform of the trace anomaly form factor.  
 $\langle \frac{\beta}{2g} \, G^{\alpha\beta}G_{\alpha\beta}\rangle$ is  the gluon condensate, a constant energy density in the vacuum, which is negative. Here we have neglected the $\sigma$ terms which are small in the nucleon.

  We should point out that the trace anomaly in the hadron has been suggested to come from the vacuum -- it is measured relative to the negative vacuum condensate~\footnote{In the calculation of
the trace anomaly (or the disconnected insertion of quark loops) in the hadron in the Euclidean path-integral, one takes the correlated insertion in the ensemble averages, i.e., $\langle O G_2\rangle - \langle O\rangle\, \langle G_2\rangle$ where $G_2$ is a hadron propagator so that the uncorrelated vacuum condensate $\langle O\rangle$ is subtracted.} -- and is identified with the bag constant in the MIT bag model~\cite{Shifman:1978by,Shuryak:1978yk,Ji:1995sv,Ji:2021pys} . The present analysis lends support to this idea. However, our derivation is based on QCD, not on models. Furthermore, the equation of state with a constant negative pressure is deduced from the correspondence between the rest energy and pressure-volume work which are derived from the QCD gravitational form factors. 
Thus, the physical picture emerges as this. Due to the conformal symmetry breaking, there is a gluon condensate in the vacuum. A hadron is formed as a bubble in this gluon condensate sea, taking a volume V and quarks and gluons are put in the volume like air molecules inside a bubble. The depletion of the gluon condensate in the hadron causes energy which is the trace anomaly matrix element of the hadron. The bubble is in equilibrium due to the balance between the negative pressure from the trace anomaly energy and the positive pressure due to the quark energy and the glue field energy inside the hadron as in Eq.~(\ref{equilibrium}). 

What's learned further in the energy-equilibrium correspondence as revealed in the GFF is that the 
trace anomaly matrix element has a linear volume dependence. This not only verifies the suggestion about the origin of the trace anomaly energy, it demonstrates that it yields a constant negative pressure and is thus the source of confinement. This is consistent with
the finding that the large Wilson loop with a spatial distance $r$ in the presence of the trace anomaly in the quenched approximation gives the potential between the infinitely heavy quarks in the form of 
$V(r) + r \frac{dV(r)}{dr}$~\cite{Dosch:1995fz,Rothe:1995hu}. Using a lattice calculation of the glue part of the trace anomaly matrix element in the charmonium~\cite{Sun:2020pda} and assuming a linear potential between the charm quarks, the deduced string tension agrees very well~\cite{Liu:2021gco} with that in the Cornell potential model used to fit the charmonium spectrum~\cite{Mateu:2018zym}.

\section{Vortices in Type II Superconductor} \label{vortices}

\begin{figure}[htbp]     \centering
{\includegraphics[width=0.4\hsize]{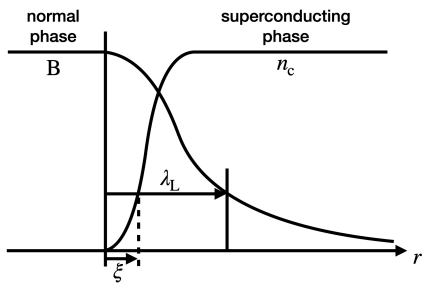}
 }
\caption{Illustration to depict a vortex in the type II superconductors between the normal phase
where the external magnetic field $B$ in the core extends out with the London penetration depth $\lambda_L$ and the superconducting phase where $n_c$ is the superconducting electron density and $\xi$ is the coherence length. \label{vortex}
}
 \end{figure}

    There is a close analogy between the hadrons and the vortices in type II superconductors as far as their energetics are concerned.  External magnetic fields penetrate through the core region and beyond with a London penetration depth $\lambda_L$ in a type II superconductor in the vortex phase is illustrated in Fig.~\ref{vortex}
in the radial direction. $n_c$ is the local density of the superconducting electrons and the coherence length $\xi$ is the
characteristic exponent of the density variations of the superconducting components. Type II is the case when the Ginzburg-Landau parameter $\kappa = \frac{\lambda_L}{\xi} > \frac{1}{\sqrt{2}}$.  

Type II superconductors can be described by the Ginzburg-Landau equation which solves the superconducting electron wavefunction $\psi(r)$ with $n_c = |\psi(r)|^2$ being the local density. Here we shall dwell on the energetics, in particular on the origin of various contributions to the energy of the core vortex~\cite{Clem1975}. There are several contributions to the Ginzburg-Landau free energy of a vortex relative to that of the Meissner state,
\begin{equation}  \label{F_energy}
F = F_B + F_{sc} + F_c,
\end{equation}
where $F_B$ and $F_{sc}$ are the magnetic field energy and the energy of supercurrents
\begin{eqnarray}
F_B &=& \frac{1}{2\mu_0} \int dv\, B^2, \label{FB}\\
F_{sc} &=& \frac{\mu_0}{2} \int dv \, \lambda_L^2 \vec{J_s} \cdot \vec{J_s},  \label{Fsc}
\end{eqnarray}
and $F_c$ is the cost of energy to deplete the Cooper pair condensate by the critical magnetic field $H_{c1}$
\begin{equation}  \label{Fc}
F_c = \frac{\kappa \phi_0\, H_{c1}}{8\pi}  \int dl \rho' d\rho' \,   (1 - n_c^2)^2 
\end{equation}
where $\rho' = \rho/\lambda_L$, $\phi_0 = hc/2e$ is the flux quanta, and $H_{c1}$ is the critical magnetic field between
the Meissner and the vortex state. $n_c$ is the normalized superconducting electron density. We see that the free energy decomposition is a close analogy to that of the hadron rest energy. The magnetic field energy $F_B$  Eq.~(\ref{FB}) corresponds to the chromo-electric and -magnetic field energies in Eq.~(\ref{H_g}). The supercurrent energy
in Eq.~(\ref{Fsc}) corresponds to the quark energy in Eq.~(\ref{H_q}). The $F_c$ in Eq.~(\ref{Fc}) is the cost of energy to 
break or reduce the superconducting pairs in the vortex region. This is analogous to the trace anomaly matrix element which is the  cost of energy to deplete the glue condensate in the QCD vacuum inside the hadron. There is a quark $\sigma$ term in Eq.~(\ref{H_m}) which appears to be missing in the free energy in Eq.~(\ref{F_energy}). This is because the matrix element of $ \bar{\psi}\psi$ in the non-relativistic limit is the same as that of $ \bar{\psi}\gamma_0\psi$
which is the fermion number. Thus, $\langle H_m\rangle$ in the condensed matter just measures the total electron mass,
which is a constant and can be factored out from the problem. 

A variational approach is carried out~\cite{Clem1975} where the the wavefunction is assumed to have the form 
$\Psi(\rho, \phi) = f(\rho) e^{-i\phi}$ where $f(\rho) = \frac{\rho}{\sqrt{\rho^2 + R^2}}$. $\rho$ is the cylindrical radial coordinate
and $R$ is the variational parameter for the core radius. This defines the density in Eq.~(\ref{Fc}) as $n_c = |\Psi|^2 = f(\rho)^2$.
The free energy per unit length per vortex line in the unit of $\phi_0 H_c/(2\sqrt{2}\pi)$ is obtained
\begin{equation}
\frac{F}{l\,\phi_0 H_c/(2\sqrt{2}\pi)}= \frac{1}{8} \kappa R'^2 + \frac{1}{8\kappa} + \frac{K_0(R')}{2\kappa\,K_1(R') R'},
\end{equation}
where $R' = R/\lambda_L$ and $K_0$ and $K_1$ are the modified Bessel functions of the second kind. 
The first term is due to the cost of removing the condensation energy in Eq.~(\ref{Fc}) which is proportional to the area. The second and third terms are from Eqs.~(\ref{FB}) and (\ref{Fsc}) and they are
dominated by the $1/R'$ behavior at short distance. From the equilibrium condition of the variation with respect to the area $-\frac{dF}{dA}A = 0$,
we see that the potential energy in Eq.~(\ref{Fc}) gives a constant negative two-dimensional pressure which is balanced by
the positive pressures from the energies of the magnetic field and the suppercurrent. 

The scripts for the confinement of hadrons and vortices in the type II superconductors are basically the same. Both have condensates from symmetry breaking. The gluon condensate is due to 
the conformal (scale) symmetry breaking and the Cooper pair condensate is due to the gauge symmetry breaking.  When the condensates are repleted to make room for quarks and glue field in a hadron in QCD and supercurrents and magnetic field in a vortex in QED, they provide constant negative pressures for confining the systems.   

There are many facets for color confinement in QCD~\cite{Greensite:2003bk,Shifman:2010jp,Brodsky:2014yha}. In this
manuscript, we consider the role of the trace anomaly matrix element, a color-singlet, in the context of the volume confinement in light hadrons and radial confinement in heavy quarkoniums.

\section{Cosmological Constant}  \label{cosmological_constant}
 
   The trace anomaly matrix element also has certain similarity to the cosmological constant. For the purpose of obtaining a static Universe, Einstein introduced a cosmological constant $\Lambda$ in his equation of general relativity~\cite{Einstein:1917ce} 
\begin{equation} \label{Einstein-eq}
R^{\mu\nu} - \frac{1}{2} R\, g^{\mu\nu} - \Lambda\, g^{\mu\nu}= 8\pi G\, T^{\mu\nu} ,
\end{equation}
where $R^{\mu\nu}$ is the Ricci curvature tensor and $R$ is the scalar curvature. $G$ is Newton's constant and
the source $T^{\mu\nu}$ is the energy-momentum tensor in general relativity. The positive constant $\Lambda$ is introduced as a  $g^{\mu\nu}$ term, so that it balances the gravitational pull from a static uniform matter density $\rho$. Einstein
found the solution of $\Lambda$ to be~\cite{Einstein:1917ce,ORaifeartaigh:2017uct}
\begin{equation}
\Lambda = 4\pi G \rho.
\end{equation}
This can be seen from the Friedmann equation for the  Friedmann-Robertson-Walker scale parameter $a(t)$
\begin{equation}   \label{Friedmann}
\frac{\ddot a}{a} = - \frac{4\pi G}{3} (\rho + \rho_{\Lambda}+ 3 (P +  P_{\Lambda})),
\end{equation}
where the energy density $\rho_{\Lambda} = \Lambda/8\pi G$ and the pressure density $P_{\Lambda} = - \Lambda/8\pi G$ have opposite signs due to the
metric $g^{\mu\nu}$.  In a matter-dominated Universe, Einstein's solution is consistent with ${\ddot a}/a = 0$.
The cosmological constant was introduced as an extra source of energy and pressure from the EMT of matter and radiation. They balance out the matter density in Einstein's static Universe. On the other hand, the trace anomaly is the consequence of renormalization in the quantum field theory. It is a quantum correction to the classical EMT whose matrix element produces a pressure to balance from those 
from the quark (matter) and gluons (radiation) for equilibrium. Both the cosmological constant and the trace anomaly matrix element are associated with the metric term, the former is in the equation of motion and the latter in the QCD EMT. As a consequence, both give positive energies and negative pressures. It is in this sense that the trace anomaly matrix element of the hadron behaves somewhat like the `hadron cosmological constant'~\cite{Liu:2021gco}.

However,  the cosmological constant works differently from the confinement mechanism in hadrons and type II superconductors, even though all share negative pressures. Contrary to gauge theories, the source in the equation
of motion in general relativity (i.e., Eq.~(\ref{Einstein-eq})) is the energy-momentum tensor. As such, the pressure from $T^{ii}$ and $g^{ii} \Lambda$ also gravitate and the negative pressure, like negative mass, bestows anti-gravity. This is contrary to the gauge theories where the negative pressure is attractive in nature. Therefore, if $3P_{\Lambda}$ from the cosmological constant (dark energy) is negative enough, i.e., $\rho + \rho_{\Lambda}+ 3 (P +  P_{\Lambda}) < 0$ in Eg.~(\ref{Friedmann}), it produce a repulsive effect so that the Universe experiences an accelerated expansion as has been observed. 

It is not clear where the cosmological constant is from. It could be some form of the vacuum energy~\cite{Zeldovich:1967gd,Weinberg:1988cp} or due to the modified gravity~\cite{Weinberg:1988cp}.
There is a suggestion that it is a condensate inside the hadrons in an attempt to explain its smallness ~\cite{Brodsky:2008xu,Brodsky:2009zd}. It might be interesting to speculate about its possible quantum origin. Imaging that the expanding Universe is embedded in a true vacuum where there is a condensate with a negative $\Lambda$, i.e. a gravitational quantum anomaly as an extra EMT term $T^{\mu\nu}_{\rm \Lambda}$ in the true vacuum with $\langle 0| T^{\mu\nu}_{\rm  \Lambda} |0\rangle = - \Lambda g^{\mu\nu}$. Like the hadrons and vortices in type II superconductors, this gives inside the Universe a positive 
$\Lambda$ in Einstein's equation and the $T^{\mu\nu}$ on the right-hand-sdie of Eq.~(\ref{Einstein-eq}) is the EMT from ordinary matter and radiation. The energy and pressure from this positive $\Lambda$ inside the Universe almost balance out those from the ordinary matter/radiation and their vacuum condensates in the Universe and, consequently, lead to an acceleration in Eq.~(\ref{Friedmann}). This scenario would put the confinement of hadrons and vortices and the accelerated expansion of the Universe on the same footing in the sense that they are the consequence of the respective condensates in their true vacuua.

\section{Pion mass and trace anomaly form factor} \label{pion}

        As pointed out in Eq.~(\ref{invariant_M}), the pion mass can be obtained from the trace of the EMT
\begin{equation} \label{pion_m}
m_{\pi}= \frac{\langle \pi | \int d^3  \vec{x}\, \gamma \big [\frac{\beta(g)}{2g} G^{\alpha\beta} G_{\alpha\beta} + 
\sum_f \gamma_m (g)\, m_f  \bar{\psi}_f \psi_f \big ] |\pi\rangle}
{\langle \pi|\pi\rangle}  + \frac{\langle \pi | \int d^3  \vec{x}\, \gamma \sum_f m_f  \bar{\psi}_f \psi_f  |\pi\rangle} 
{\langle \pi|\pi\rangle}
\end{equation}
Ignoring the strangeness (it is found to be negligibly small for light pions in a lattice calculation~\cite{Yang:2014xsa}), the second term (sigma term) gives half of the pion mass. This can be proven
from the \mbox{Gell-Mann-Oakes-Renner} relation $f_{\pi}^2 m_{\pi}^2 = - ({\rm m_u} \langle \bar{u}u \rangle + {\rm m_d} \langle \bar{d}d \rangle)$ and the Feynman-Hellman theorem ${\rm m_u} \partial m_{\pi}/\partial {\rm m_u} + {\rm m_d} \partial m_{\pi}/\partial {\rm m_d}
= {\rm m_u} \langle \pi |\bar{u}u| \pi \rangle + {\rm m_d} \langle \pi |\bar{d}d |\pi\rangle$~\cite{Ji:1995sv,Yang:2014xsa}. Furthermore, it is
proportion to $\sqrt{m_q}$ for the SU(2) case where $m_q = m_u = m_d$. A question arises as to why the trace anomaly matrix element in the first term of Eq.~(\ref{pion_m}) should also decrease with the quark mass as $\sqrt{m_q}$. There is no obvious known symmetry reason that dictates the behavior of the trace anomaly matrix element in the vicinity of the chiral limit. Does it mean that the size (e.g., root-mean-square radius) from the effective volume defined in Eq.~(\ref{volume}) vanishes at the chiral limit? Also, in the analysis of the pion rest energy in a lattice calculation~\cite{Yang:2014xsa}, all the terms are positive and approach zero in the chiral limit. There are no cancellations among these contributions.

\begin{figure}[hbtp]    
 \centering
{\includegraphics[width=0.5\hsize]{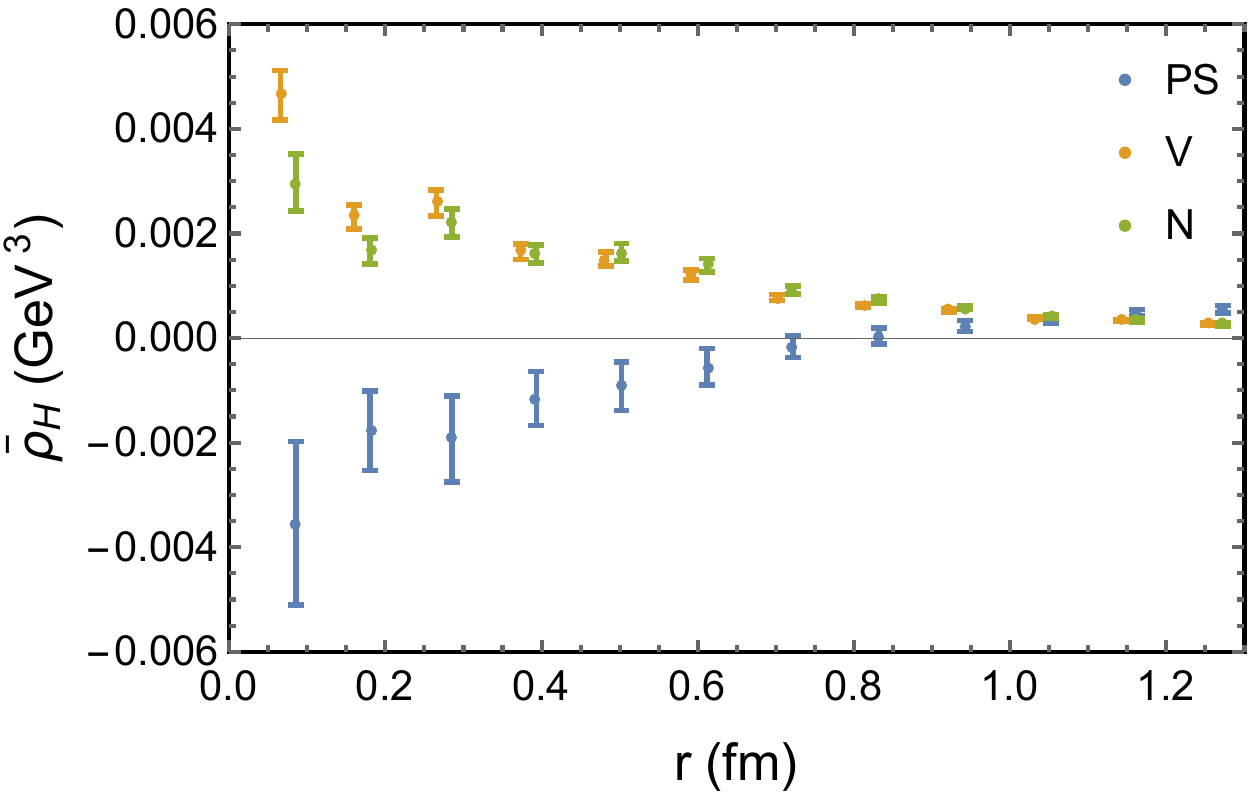}
 }
\caption{The distribution of the glue part of the trace anomaly $\bar{\rho}_H$ in the nucleon, $\rho$ and $\pi$ as a function of the distance between the glue operator and the sink position of the respective hadron propagator. This is from Ref.~\cite{He:2021bof}.
 \label{pion-ta}
}
  \end{figure}

   In light of this puzzle, a lattice calculation has been carried out to examine the spatial distribution $\bar{\rho}_H$  in the nucleon, $\rho$ and    pion~\cite{He:2021bof}. The spatial coordinate is between the glue part of the trace anomaly operator and the sink position
of the interpolation field of the hadron so that they may in some way related to the Fourier transform of the form factor of the glue operator in the Breit frame. The results of $\bar{\rho}_H$ are plotted in Fig.~\ref{pion-ta}. We see that the density distributions for the nucleon and the $\rho$ are monotonic as in the proton electric and axial charge distributions. However, the distribution for the pion is unusual. When the quark mass is small, the distribution changes sign such that the integral of the distribution vanishes at the chiral limit. It is verified~\cite{He2022b} that the  matrix element $\langle \pi | \int d^3  \vec{x}\, \gamma \frac{\beta(g)}{2g} G^{\alpha\beta} G_{\alpha\beta} |\pi\rangle/
\langle \pi|\pi\rangle $ is proportional to $\sqrt{m_q}$, as expected, in the partially-quenched calculation with different valence quark masses. This answers the question we put forth above. As far as the trace anomaly matrix element is concerned, the pion still has a finite size, i.e., $\langle r^2\rangle|_{\rm{ta}} > 0$, even though the effective volume defined in Eq.~(\ref{volume}) diminishes as the chiral limit is approached due to the spatial cancellation of the glue trace anomaly density.  This is achieved by modifying the structure of the vacuum condensate -- making
the glue condensate more negative than that in the vacuum in the inner core of the pion and more positive than that of the vacuum in the outer shell so that it takes no energy to create a pion with massless quarks. Even though this explains how the
trace anomaly matrix element approaches zero at the chiral limit, it does not explain why it should happen this way. 

This apparently is a feature of the state. It happen to the Goldstone boson state, but not in the nucleon and the $\rho$. But it is also due to the kind of the probing operator involved. It happens to the trace anomaly distribution, but the electric form factor of the pion is not unusual and is similar to that of the proton~\cite{Wang:2020nbf}. Thus, the structure of the trace anomaly distribution in the pion appears to be the result of the combined effects of the operator and the state. 
Glue condensate is the consequence of conformal symmetry breaking and chiral symmetry breaking leads to the Gell-Mann-Oakes-Renner relation. Even though there is no dynamical or symmetry explanation at this stage, it nevertheless suggests that the conformal (scale) symmetry breaking and chiral symmetry breaking are intricately coupled in QCD in the pion. There are efforts to look for conformal windows with multiflavor simulations~\cite{DelDebbio:2010zz}. One could examine the relation between chiral symmetry and conformal symmetry in these studies. Also, using it as an indicator, one could calculate it in nuclei to see if the conformal symmetry is partially restored.

Since there is a node in the spatial distribution of the trace anomaly in Fig.~\ref{pion-ta}, one expects that the glue trace anomaly form factor of the pion will also change sign. The preliminary lattice calculation shows that it indeed does so~\cite{Wang2023}. It would be interesting to detect this experimentally, such as via the $J/\Psi$ production at the threshold of photoreaction~\cite{Kharzeev:1995ij,Hatta:2018ina,Duran:2022xag}. Another way to access the trace anomaly form factor is through the GFF which are the moments of GPD. We shall first look at the nucleon case. The form factor of the EMT trace, i.e., the mass form factor $G_m(Q^2)$, is defined in
\begin{equation}
\langle P'|T_{\mu}^{\mu}|P\rangle/2M = (\frac{M^2}{E_P E_{P'}})^{1/2} \bar{u}(P') u(P) G_m(Q^2),
\end{equation}
where $T_{\mu}^{\mu}$ is given in Eqs.~(\ref{trace}) and (\ref{trace_q,g}). $G_m$ is made up
of two parts
\begin{equation} \label{G_m}
G_m (Q^2) = G_{\rm{ta}} (Q^2)+ G_{\sigma} (Q^2),
\end{equation}
where $G_{\rm{ta}}$ is the form factor for the trace anomaly and $G_{\sigma}$ the form factor  for the sigma term. 

From the GFF, the form factor of the trace is~\cite{Ji:2021mtz} 
\begin{eqnarray}  \label{G_trace}
G_{\rm{trace}}(Q^2) =  A(Q^2) M - B(Q^2)\frac{Q^2}{4M} + 3 D (Q^2)\frac{Q^2}{M} 
+ 4 \bar{C}(Q^2)M,
\end{eqnarray}
where $A(Q^2), B(Q^2), D(q^2)$ and $\bar{C}(Q^2)$ contain both the quark and gluon components, e.g., $A(Q^2) = A_q(Q^2) + A_g(Q^2)$. Extending the derivation of the forward $\bar{C}_q (0)+ \bar{C}_g (0)$ in Eq.~(\ref{barC}) to
finite $Q^2$, we obtain
\begin{eqnarray}  \label{barCQ2}
\bar{C}(Q^2) = \frac{1}{4}  G_m (Q^2) - \frac{1}{4} \big [( A(Q^2) M - B(Q^2)\frac{Q^2}{4M} 
+ 3 D(Q^2)\frac{Q^2}{M}\big ].
\end{eqnarray}
Substituting this in Eq.~(\ref{G_trace}), one arrives at
\begin{equation}
G_{\rm{trace}}(Q^2) \equiv G_m(Q^2).
\end{equation}
Thus, one obtains the same expression for $G_m(Q^2)$ from the GFF as expected.
Since $\bar{C}(Q^2) = 0$ due to the energy and momentum conservation, i.e.,
$q_{\mu} T^{\mu\nu} = 0$, the trace anomaly form factor $G_{\rm{ta}}$ can be obtained
from Eqs.~(\ref{G_m}) and (\ref{barCQ2})
\begin{equation}  \label{G_ta}
G_{\rm{ta}}(Q^2) =  \big [( A(Q^2) M - B(Q^2) \frac{Q^2}{4M} + 3 D (Q^2)\frac{Q^2}{M} \big ] - G_{\sigma} (Q^2) 
\end{equation}
One can obtain $G_m(Q^2)$ by simply dropping the $\bar{C}(Q^2)$ from the GFF in Eq.~(\ref{G_trace})~\cite{Ji:2021mtz}  since it is zero and obtains the expression $ A(Q^2) M - B(Q^2) \frac{Q^2}{4M} + 3 D (Q^2)\frac{Q^2}{M}$ which turns out to be
equal to $G_m(Q^2)$ from Eq.~(\ref{barCQ2}) due to energy-momentum conservation which requires $\bar{C}(Q^2) = 0$.

The $A(Q^2), B(Q^2)$ and $D(Q^2)$ are related to the second moments of the generalized parton distribution (GPD) 
$H(x,\xi,Q^2)$ and $E(x,\xi,Q^2)$~\cite{Ji:1996nm,Diehl:2003ny}, i.e., 
\begin{eqnarray}  \label{GPD}
\int_{-1}^1 dx\, x\, H(x,\xi,Q^2) \!\!\!&=&\!\!\! A (Q^2)+   \xi^2 D(Q^2) \nonumber \\
\int_{-1}^1 dx\, x\, E(x,\xi,Q^2) \!\!\!&=&\!\!\!  B(Q^2)  -  \xi^2 D(Q^2)
\end{eqnarray}
where $\xi$ is the skewness. In the nucleon, the forward $B(0) = 0$ from the momentum and angular momentum sum rules~\cite{Ji:1997pf} and the realization that there is no gravito-magnetic moment~\cite{Brodsky:2000ii}.
A quenched lattice calculation~\cite{Deka:2013zha} suggests that $B_q(Q^2)$ and $B_g(Q^2)$ at finite $Q^2$ have different sign and both of them are fairly flat for $Q^2 < 2\, {\rm GeV}^2$. They cancel at $Q^2 = 0$ so that the total $B(Q^2)$ is expected to be small in this region of $Q^2$. To the extend that $B(Q^2)$ is negligible, the GFF $A(Q^2)$ and $D(Q^2)$ can be obtained from the GPD in Eq.~(\ref{GPD}). Together with $G_{\sigma}(Q^2)$ from lattice calculations, the trace anomaly form factor 
$G_{\rm{ta}}(Q^2)$ in Eq.~(\ref{G_ta}), a twist-four matrix element can be obtained from the matrix elements of twist-two and twist-three operators.

In the case of pion, there is no $B(Q^2)$, the trace anomaly form factor can be similarly expressed in terms of the GFF and the 
sigma term
\begin{equation}  \label{G_ta-pion}
G_{\rm{ta}}(Q^2)_{\rm{pion}} =  \big [( A(Q^2) m_{\pi} + 3 D (Q^2)\frac{Q^2}{m_{\pi}} \big ] - G_{\sigma} (Q^2) 
\end{equation}
so that it can be obtained from the  $H$ and $E$ of the pion GPD and the lattice calculation of the pion sigma term.

\section{Summary}   \label{summary}

We derive the pressure $\bar{C}$ in th scale-invariant form and showed that the rest energy from the Hamiltonian and the gravitational form factors (GFF) yield the same decomposition when $\bar{C}$ is
included. We found that components of the rest energy of hadrons have a one-to-one correspondence with those of the free energy of the vortices in type II superconductors. 
Even the scripts for their confinement are basically the same -- the respective vacuum condensates from the symmetry breaking are depleted to accommodate the hadrons and the vortices with positive energies which are proportional to the volume ({\it N.B.} this is inferred from the energy-equilibrium correspondence as derived from the GFF) and area. This results in a constant negative confining pressure to balance the positive pressures from the fermion and gauge field energies in both cases. Heavy quarkoniums have a similar picture, where the glue part of the trace anomaly matrix element is proportional to the volume of a flux tube. With a constant transverse electric field distribution in the cross section of the tube, found in lattice calculations~\cite{Bali:1997am,Baker:2018mhw}, the potential energy is linear in the distance between the heavy quark and antiquark, leading to a linear confinement. 
     
      We have also drawn an analogy between the trace anomaly matrix element and the cosmological constant as a metric term in the general relativity equation, which provides a constant negative pressure to balance the gravitational pull of the matter in Einstein's static Universe. However, there is a fundamental difference between the gauge theories and the general relativity. In general relativity, the source of the equation of motion is the EMT and the cosmological constant gives an extra source of energy and pressure. A negative pressure from the cosmological constant (dark energy) anti-gravitates. This gives a repulsive effect which is opposite to the effect of a negative pressure in gauge theories. When it is more negative than the matter/radiation densities and their condensates, the expansion of the Universe will accelerate according to the Friedmann's equation in Eq.~(\ref{Friedmann}).

   The role of the glue part of the trace anomaly matrix element is illustrated further in regard to the pion mass. An intriguing structure of the trace anomaly in the pion is found to be responsible for answering the question regarding the trace anomaly part of the pion mass. As the quark mass approaches zero, the spatial distribution of the glue part of the trace anomaly changes sign so that the trace anomaly matrix element is proportional to $\sqrt{m_q}$. This change of sign is reflected in the pion trace anomaly form factor in lattice calculation~\cite{Wang2023} and should be verified experimentally, such as the $J/\Psi$ production at the threshold of photoreaction~\cite{Kharzeev:1995ij,Hatta:2018ina,Duran:2022xag} and, alternatively, with the GFF from GPD and the sigma term. This suggests that the conformal (scale) symmetry breaking in the pion is somehow linked to the chiral symmetry breaking. 
           
       Since the trace anomaly matrix element of the hadron is an indicator of confinement at zero temperature, it could somehow serve as the confinement-deconfinement order parameter. This opens up a number of issues on the nature of phase transitions that can be studied on the lattice. Through the study of the spectral density in terms of the overlap Dirac eigenvalues, there is an evidence of a phase above the crossover temperature that displays infrared scale invariance~\cite{Alexandru:2019gdm}. It would be useful to find out what bearing it may have on the glue condensate as a function of the temperature and chemical potential.

\section{Acknowledgment}
The author is indebted to P. Boyle, S. Brodsky, V. Burkert, M. Chanowitz, S. Das, T. Draper, A. Dymarsky, W. Gannon, I. Horv\'{a}th, T. Hatsuda, F. He, Y. Hatta, X. Ji, D.E. Kharzeev, D. Lin, C. Lorc\'{e},  A. Metz, \mbox{Z. Meziani,} G. Murthy, J.C. Peng, M. Peshkin, A. Shapere, O.V. Teryaev, B. Wang, Y.B. Yang, and F. Yuan for fruitful discussions. He also thanks T.J. Hou for providing the CT18 data and B. Wang for help with the figures. This work is partially supported by the U.S. DOE Grant No. DE-SC0013065 and No.\ DE-AC05-06OR23177 which is within the framework of the TMD Topical Collaboration.
This research used resources of the Oak Ridge Leadership Computing Facility at the Oak Ridge National Laboratory, which is supported by the Office of Science of the U.S. Department of Energy under Contract No.\ DE-AC05-00OR22725. This work used Stampede time under the Extreme Science and Engineering Discovery Environment (XSEDE), which is supported by National Science Foundation Grant No. ACI-1053575.
We also thank the National Energy Research Scientific Computing Center (NERSC) for providing HPC resources that have contributed to the research results reported within this paper.
We acknowledge the facilities of the USQCD Collaboration used for this research in part, which are funded by the Office of Science of the U.S. Department of Energy.


\end{document}